\documentclass{aa}  
\usepackage[]{natbib}
\usepackage{graphicx}
\usepackage{txfonts}
\begin{document}

   \title{Limits on the neutrino magnetic dipole moment\\ 
from the luminosity function of hot white dwarfs}

   \author{Marcelo Miguel Miller Bertolami
          \inst{1,2}}

   \institute{Max-Planck-Institut f\"ur Astrophysik, Karl-Schwarzschild-Str. 1, 8574, Garching, Germany.\\
              \email{marcelo@MPA-Garching.MPG.DE}
         \and
             Instituto de Astrof\'isica de La Plata, UNLP-CONICET, Paseo del Bosque s/n, 1900 La Plata, Argentina  (on leave of absence).\\
             \email{mmiller@fcaglp.unlp.edu.ar}
             }

   \date{}

\abstract{Recent determinations of the white dwarf luminosity function (WDLF)
  from very large surveys have extended our knowledge of the WDLF to very high
  luminosities. This, together with the availability of new full evolutionary
  white dwarf models that are reliable at high luminosities, have opened the
  possibility of testing particle emission in the core of very hot white
  dwarfs, where neutrino processes are dominant.}{We use the available WDLFs
  from the Sloan Digital Sky Survey and the SuperCOSMOS Sky Survey to
  constrain the value of the neutrino magnetic dipole moment ($\mu_\nu$).}{We
  used a state-of-the-art stellar evolution code to compute a grid of white
  dwarf cooling sequences under the assumptions of different values of
  $\mu_\nu$. Then we constructed theoretical WDLFs for different values of
  $\mu_\nu$ and performed a $\chi^2$-test to derive constraints on the value
  of $\mu_\nu$.}{We find that the WDLFs derived from the Sloan Digital Sky
  Survey and the SuperCOSMOS Sky Survey do not yield consistent results. The
  discrepancy between the two WDLFs suggests that the uncertainties are
  significantly underestimated. Consequently, we constructed a unified WDLF by
  averaging the SDSS and SSS and estimated the uncertainties by taking into
  account the differences between the WDLF at each magnitude bin. Then we
  compared all WDLFs with theoretical WDLFs.  Comparison between theoretical
  WDLFs and both the SDSS and the averaged WDLF indicates that $\mu_\nu$
  should be $\mu_\nu<5\times 10^{-12}\, e\hbar/(2m_e c)$.  In particular, a
  $\chi^2$-test on the averaged WDLF suggests that observations of the disk
  WDLF exclude values of $\mu_\nu>5\times 10^{-12}e\hbar/(2m_e c)$ at more
  than a 95\% confidence level, even when conservative estimates of the
  uncertainties are adopted. This is close to the best available constraints
  on $\mu_\nu$ from the physics of globular clusters.}{Our study shows that
  modern WDLFs, which extend to the high-luminosity regime, are an excellent
  tool for constraining the emission of particles in the core of hot white
  dwarfs. However, discrepancies between different WDLFs suggest there might
  be some relevant unaccounted systematic errors. A larger set of completely
  independent WDLFs, as well as more detailed studies of the theoretical WDLFs
  and their own uncertainties, is desirable to explore the systematic
  uncertainties behind this constraint. Once this is done, we believe the
  Galactic disk WDLF will offer constraints on the magnetic dipole moment of
  the neutrino similar to the best available constraints obtainable from
  globular clusters.}


\titlerunning{Limits on the neutrino dipole moment from the WDLF}
\authorrunning{Marcelo M. Miller Bertolami} 

   \maketitle
%

\section{Introduction}
In the Standard Model of particle physics, neutrinos are massless and only
have weak force interactions.  However, the confirmation of neutrino
oscillations by many experiments \citep{2007fnpa.book.....G} imply that they
must be massive and mixed. In fact, the neutrino is the only particle, up to
now, that really exhibits properties beyond the Standard Model. Consequently,
the Standard Model must be extended to cope with neutrino masses.

In many extensions of the Standard Model
neutrinos acquire electromagnetic properties (see
\citealt{2012arXiv1207.3980B} for a review), which makes studying the
electromagnetic properties of neutrinos a basic tool for investigating the
physics beyond the Standard Model. The idea that neutrinos could have
a magnetic dipole moment ($\mu_\nu$) is as old as the idea of the
neutrinos themselves. Pauli discussed the possibility that
the neutrino might have an intrinsic magnetic dipole moment in the
same letter in which he proposed the very existence of the particle
\citep{1930Pauli}.

Even the very feeble weak interaction between the neutrino and the electron
has a huge impact on the hot dense plasmas found in astrophysical environments
\citep{1964ApJS....9..201F}. Indeed, the dominant energy-loss mechanism in hot
white dwarfs occurs through neutrino emission
\citep{1975ApJ...200..306L}. Effective coupling in a plasma occurs between the
neutrino and the electromagnetic field (photons) by means of the ambient
electrons of the medium. The most interesting process is the plasma process
($\gamma \rightarrow \nu\bar{\nu}$) in which a photon decays into a
neutrino/anti-neutrino pair. This process becomes kinematically possible
because in a plasma the electromagnetic field acquires a dispersion relation,
$(\omega^2-c^2\vec{k}^2)\hbar^2=\Pi^2>0$, which roughly amounts to an
effective photon mass. Then, plasmons\footnote{The term plasmon refers to
  the electromagnetic excitations of the medium.}  with $\Pi>2 m_\nu c^2$ can
decay into a neutrino/anti-neutrino pair. As mentioned above, this is only
possible within the Standard Model, because of the indirect coupling of the
neutrino and the photon that is mediated by the ambient electrons.

If a direct interaction between neutrinos and photons is allowed, for example,
by a neutrino magnetic dipole moment or a neutrino millicharge, then the
neutrino emission through plasmon decays will be enhanced. This will lead to
significant observable consequences for the stellar evolution theory (see
\citealt{1996slfp.book.....R, 2000PhR...333..593R, 2012arXiv1201.1637R} for
very detailed reviews and \citealt{2009ApJ...696..608H} for a discussion of
the impact on massive stars). \citet{1990ApJ...365..559R} and then
\citet{1992A&A...264..536R} showed that the properties of red giants from the
color-magnitude diagram (CMD) of Galactic globular clusters implied
$\mu_\nu\lesssim 3\times 10^{-12}\mu_{\rm B}$, where $\mu_{\rm B}=e\hbar/(2m_e
c)$ is the Bohr magneton. More recently, \citet{2013EPJWC..4302004V,
  2013arXiv1308.4627V} analyzed the CMD of the Galactic globular cluster M5
and concluded that $\mu_\nu <4.5\times 10^{-12}\mu_{\rm B}$ at the 95\%
confidence level.

In particular a neutrino magnetic dipole moment will influence the cooling of
hot white dwarfs and lead to observable consequences in the white dwarf
luminosity function (WLDF, \citealt{1994MNRAS.266..289B}) and the rate of
period change of hydrogen-deficient pulsating white dwarfs (DOVs and DBVs,
\citealt{2000ApJ...539..372O, 2004ApJ...602L.109W}). Unfortunately, up to now
it has been impossible to determine the rate of period change generated by
secular cooling in DOVs or DBVs \citep{2011A&A...528A...5V,
  2013MNRAS.431..520C, 2013ApJ...765....5D}, although some hints have recently
been found by \citet{2011MNRAS.415.1220R}. \citet{1994MNRAS.266..289B} showed
that the early WDLF of \citet{1986ApJ...308..176F} implied that the magnetic
dipole moment is $\mu_\nu\lesssim 10^{-11}\mu_{\rm B}$.

\citet{2008ApJ...682L.109I} have shown, in the context of the strong CP
problem, that modern WDLFs offer a new possibility to learn about elementary
particle physics.  In this work we show that, with the recent determinations
of the hot end of the white dwarf luminosity function
\citep{2009A&A...508..339K, 2011MNRAS.417...93R} and with the aid of
state-of-the-art white dwarf models, it is possible to obtain constraints on
the value of $\mu_\nu$ that can compete with those derived from red giants and
globular clusters. Because astrophysical determinations are often prone to
unknown systematic uncertainties, the determination of similar constraints by
different and independent methods is desirable. To take into account possible
systematic errors in the determination of the WDLF, we relied on two
completely independent sets of WDLFs derived from the Sloan Digital Sky Survey
(SDSS) and the SuperCosmos Sky Survey (SSS). These helped us to estimate the
real uncertainties behind the observed WDLFs and derive constraints on the
value of $\mu_\nu$. Finally, we suggest work that can be carried out to
improve the use of the WDLF as a tool to constrain the electromagnetic
properties of the neutrino.

\section{Input physics and white dwarf models}
The calculations reported here were done using the {\tt LPCODE}
stellar evolutionary code \citep{2012A&A...537A..33A}. This code has
been used to study different problems related to the formation and
evolution of white dwarfs \citep{2010Natur.465..194G,
 2010ApJ...717..183R, 2013ApJ...775L..22M}.  A
description of the input physics and numerical procedures employed in
{\tt LPCODE} can be found in these works. Here we only summarize some
points of specific interest for the present work. 

For the white dwarf regime, {\tt LPCODE} takes into account the effects of
element diffusion caused by gravitational settling, and chemical and thermal
diffusion, see \cite{2003A&A...404..593A} for details.  Both latent heat
release and the release of gravitational energy resulting from carbon-oxygen
phase separation \citep{2000ApJ...528..397I, 1997ApJ...485..308I} were
included following the phase diagram of \cite{2010PhRvL.104w1101H}, see
\cite{2012A&A...537A..33A} for details of the numerical implementation. The
radiative opacities are those of OPAL \citep{1996ApJ...464..943I}.  The
conductive opacities were taken from \cite{2007ApJ...661.1094C}.  Finally, we
emphasize that recently, {\tt LPCODE} has been tested against other white
dwarf evolutionary code, and the uncertainties in the cooling ages arising from
different numerical implementations of stellar evolution equations were found
to be lower than 2$\%$ \citep{2013A&A...555A..96S}.

Including the anomalous energy loss due to the existence of a
neutrino magnetic dipole moment ($\epsilon_\nu^{\rm dip}$) is
relatively simple because it is possible to relate the
anomalous neutrino emission to the plasmon neutrino emission predicted
by the Standard Model ($\epsilon_\nu^{\rm plas}$). This scaling
relation has been computed by \citet{1994ApJ...425..222H}, who found
\begin{equation}
\epsilon_\nu^{\rm dip}=0.318\ \mu_{12}^2 \left(\frac{10
  {\rm keV}}{\hbar \omega_P}\right)^2 \frac{Q_2}{Q_3}\ \epsilon_\nu^{\rm plas},
\end{equation}
where\footnote{Here $\mu_\nu$ is the effective magnetic dipole moment defined by
$$
{\mu_\nu}^2=\sum_{i,j=1}^{3}\left(|\mu_{ij}|^2+|\epsilon_{ij}|^2\right),
$$
where $\mu_{ij}$ and $\epsilon_{ij}$ are the matrices of the magnetic and electric dipole and transition moments.}
 $\mu_{12}=\mu_\nu/(10^{-12} \mu_B)$, $\omega_P$ is the plasma
frequency and the ratio $Q_2/Q_3$ has a complex expression, but is very
close to unity. For simplicity, we approximated $Q_2/Q_3\sim 1$ because
this approximation introduces errors lower than 10\%
\citep{1994ApJ...425..222H}. It leads to
a slight underestimation of the anomalous energy loss and, thus, to
more conservative conclusions. In addition, following
\citet{1996slfp.book.....R}, we approximated the plasma frequency
by its zero temperature value
\begin{equation}
\hbar^2 {\omega_P}^2\simeq \frac{4\pi n_e e^2 \hbar^2}{m_e}\left[1+\left(\frac{\hbar}{m_e c}\right)^2 (3\pi^2n_e)^{2/3}\right]^{-1/2}, 
\end{equation}
where $n_e$ is the electron density by number and $m_e$ is the mass of
the electron.

To compute the impact of the magnetic dipole moment on the WDLF we computed a
small grid of white dwarf cooling sequences that covers the relevant mass
range under the assumption of different values of the neutrino magnetic dipole
moment. For a correct assessment of the white dwarf cooling times at the
high-luminosity end of modern WDLFs, accurate initial models are
required. This can not be achieved using artificial initial white dwarf
structures because at these luminosities white dwarf structures are still
dependent on the previous evolution.  For this reason the initial white dwarf
models adopted here were taken from \cite{2010ApJ...717..183R}, who computed
the full evolutionary calculation of the stages that lead to the formation of
DA white dwarfs ---from the zero-age main-sequence to the asymptotic giant
branch. Even with these full evolutionary models, some initial relaxation of
the models is required because the initial structures correspond to models
computed under the assumption of standard neutrino losses. Fortunately, we do
not expect large changes in the pre-white dwarf evolution
\citep{2013EPJWC..4302004V, 2013arXiv1308.4627V} because the value of
$\mu_{\nu}$ is relatively low.  The masses of the initial white dwarf models
selected for our grid were 0.52490$M_\odot$, 0.57015$M_\odot$,
0.60959$M_\odot$, 0.70511$M_\odot$, and 0.87790$M_\odot$. For each of the five
initial white dwarf models, five cooling sequences were computed under different
assumed magnetic dipole moments; $\mu_{12}$ = 0 (standard sequences), 1, 2, 5
and 10.

\begin{figure}[ht!]
\begin{center}
\includegraphics[clip, angle=0, width=8.5cm]{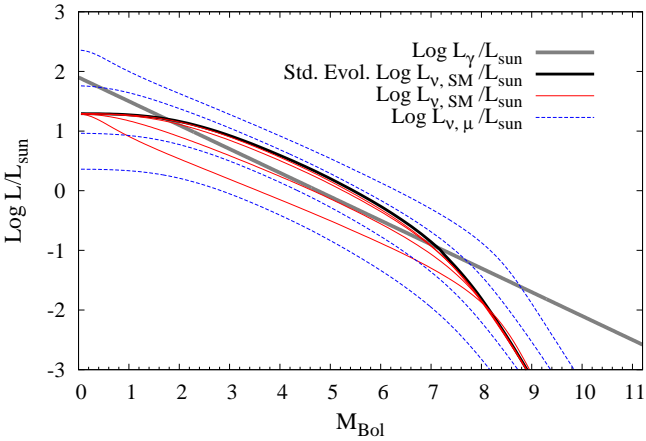} 
\caption{Total neutrino energy losses ($L_\nu$) for the 0.60959 M$_\odot$
  sequences under different assumed values of the neutrino magnetic dipole
  moment. The thick black line shows the value of the total neutrino emission
  without a magnetic dipole moment, while the thin red lines show the
  neutrino emission due to the standard model processes under the assumption
  of $\mu_{12}=1,2,5$, and 10 (from top to bottom). Dashed blue lines show the
  anomalous neutrino emission under the assumption of $\mu_{12}=1,2,5$, and 10
  (from bottom to top).}
\label{Fig:WDCooling}
\end{center}
\end{figure}

\begin{figure}[h!]
\begin{center}
\includegraphics[clip, angle=0, width=8.5cm]{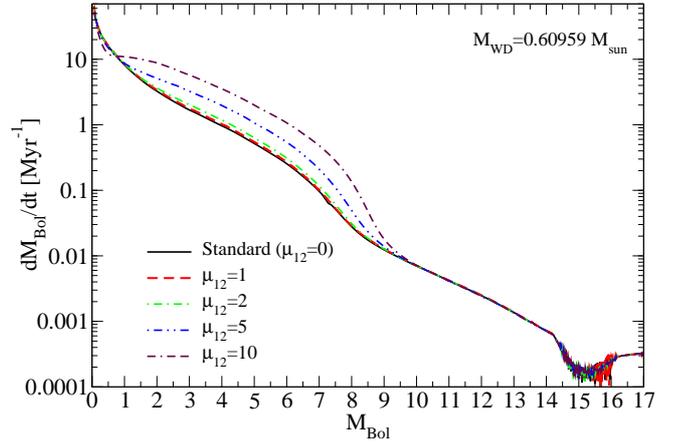} 
\caption{Cooling speeds of the 0.60959 M$_\odot$ sequences computed under
  different assumed values of the neutrino magnetic dipole moment
  ($\mu_{12}=0, 1,2,5$ and 10).}
\label{Fig:DMbolDt_Mbol}
\end{center}
\end{figure}

In Fig. \ref{Fig:WDCooling}, we show the total neutrino energy losses for a
0.60959 M$_\odot$ model under standard assumptions and under the
assumption of a neutrino magnetic dipole moment. An inspection of the emission
rates in Fig. \ref{Fig:WDCooling} shows that for $M_{\rm Bol}\lesssim 2$ the
models are still relaxing to the new value of the neutrino
emission. Consequently, our WDLFs will not be reliable at these bolometric
magnitudes. Fig. \ref{Fig:WDCooling} also shows that as the anomalous neutrino
emission (due to the value of $\mu_{12}$) is increased by increasing
$\mu_{12}$, the feedback on the thermal structure of the white dwarf leads to
lower neutrino emission through the standard channels. In
Fig. \ref{Fig:DMbolDt_Mbol} we show the overall impact of the neutrino
magnetic dipole moment on the cooling speed of the white dwarf. As expected,
after the relaxation phase at $M_{\rm Bol}\lesssim 2$, higher values of
$\mu_\nu$ lead to higher cooling speeds of the white dwarf.  Finally, it is
worth noting  (see Figs. \ref{Fig:WDCooling} and \ref{Fig:DMbolDt_Mbol})
that for all the adopted values of $\mu_{12}$ the neutrino emission becomes
negligible at $M_{\rm Bol}> 10$, and thus the shape of the WDLF will not be
affected at $M_{\rm Bol}> 10$. This sets the region where normalizations of
the WDLF can be performed to study the anomalous neutrino emission.


\section{Theoretical white dwarf luminosity functions}
\begin{figure*}[t!]
\begin{center}
\includegraphics[clip, angle=0, width=12cm]{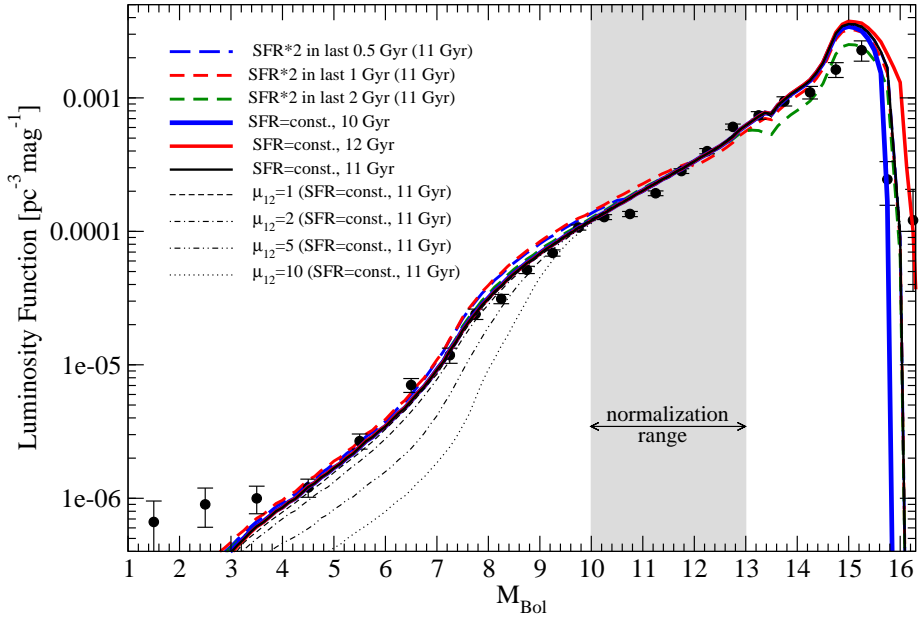} 
\caption{WDLFs computed under the assumption of different stellar formation
  rates and ages compared with the WDLF derived from the SDSS
  \citep{2006AJ....131..571H, 2009A&A...508..339K}. See text for discussion.}
\label{Fig:WDLF_test}
\end{center}
\end{figure*}
The numerical code used to construct the theoretical WDLFs is a
rewritten version of the code used in Melendez et al. (2012), which is
based on the method described by \citet{1989ApJ...341..312I}.  A
detailed explanation of the method can be found in
\cite{1989ApJ...341..312I}, and here we only describe the main points
of interest for the present study. In this approach, the number of
white dwarfs per logarithmic luminosity and volume is computed as
\begin{equation}
\frac{dn}{dl}=-\int_{M_1}^{M_2} \psi(t) \left(\frac{dN}{dM}\right)
 \left(\frac{\partial t_c}{\partial l}\right)_m dM,
\label{eq:dndl}
\end{equation}
where $\psi(t)$ is the Galactic stellar formation rate at time $t$,
$N(M)$ is the initial mass function and $t_c(l,m)$ is the time since
the formation of a white dwarf, of mass $m$, for the star to reach a
luminosity $\log(L/L_\odot)=l$. To compute the integral in
Eq. \ref{eq:dndl} we also need the initial-final mass relation
$m(M)$, and the pre-white dwarf stellar lifetime $t_{ev}(M)$. It is
worth noting that for a given white dwarf luminosity ($l$) and mass
of the progenitor ($M$), the formation time of the star, $t$, is
obtained by solving
\begin{equation}
t+t_{ev}(M)+t_c(l,m)=T_D,
\label{eq:time}
\end{equation}
where $T_D$ is the age of the oldest computed stars. The lowest initial mass
that produces a white dwarf with luminosity $l$ at the present time
($M_1$) is obtained from Eq. \ref{eq:time} when $t=0$. The value of
$M_2$ corresponds to the largest stellar mass progenitor that produces
a white dwarf. In addition, to compute Eq.  \ref{eq:dndl} we adopted a
Salpeter initial mass function, the initial-final mass relation from
\cite{2009ApJ...692.1013S}, and the stellar lifetimes from the BaSTI
database \citep{2004ApJ...612..168P}. A constant stellar formation rate
(SFR) was assumed for the reasons explained below, unless otherwise
stated.

Owing to the uncertainties in the absolute value of $\psi(t)$ in
Eq. \ref{eq:dndl} it is necessary to rescale the theoretical WDLFs to fit the
observed WDLF in some luminosity range to compare the theoretical WDLFs with
those derived from observations of the Galactic disk
\citep{2006AJ....131..571H, 2008AJ....135....1D, 2009A&A...508..339K,
  2011MNRAS.417...93R, 2013ASPC..469...83K}. As shown in
Fig. \ref{Fig:WDLF_test}, anomalous neutrino emission only affects the bright
end of the WDLF. In particular, the WDLF remains unchanged at $M_{\rm Bol}>10$
under the assumptions of different values for the neutrino magnetic dipole
moment. In addition, as noted by \citet{2008ApJ...682L.109I}, observational
errors as well as theoretical uncertainties, such as the SFR, are lowest
around $M_{\rm Bol}\sim 12$. For this reason, we chose to normalize the
theoretical WDLFs so that they give the same number of stars per volume in the
range $10 \leq M_{\rm Bol} \leq 13 $. This is, we set
\begin{equation}
\sum_{M_{\rm Bol}^i\in (M_{\rm Bol}^1,M_{\rm Bol}^2)} n(M_{\rm Bol})^i  \Delta M_{\rm Bol}=
\int_{M_{\rm Bol}^1}^{M_{\rm Bol}^2} \frac{dn}{dl}  dM_{\rm Bol}.
\label{eq:norm2}
\end{equation}
When dealing with the data of \citet{2006AJ....131..571H},
  \citet{2011MNRAS.417...93R}, and \citet{2009A&A...508..339K}, we took $M_{\rm
  Bol}^1=10$ and $M_{\rm Bol}^2=13$.
We emphasize that the theoretical WDLFs need to be normalized to fit
\emph{each} observed WDLF. To take advantage of the fact that modern
WDLFs extend to very high luminosities, where the neutrino emissivities are
most important, and taking into account that the theoretical white dwarf
models adopted in this work are are still not relaxed at $ M_{\rm Bol}\lesssim
2.5$, we compared the theoretical and observationally derived WDLF within the
range\footnote{Throughout this work the assumed relationship between the
  bolometric magnitude and the luminosity of the star is adopted consistently
  with the observational SDSS data, i.e. $M_{\rm Bol}=-2.5
  \log(L/L_\odot)+4.75$.} $3\leq M_{\rm Bol}\leq 9$.

In addition, in Fig. \ref{Fig:WDLF_test} we show that the bright end of the
WDLF is almost independent of the SFR or the age of the disk, because their
main effects are absorbed in the normalization procedure, in agreement with
the results of \citet{2008ApJ...682L.109I}. In particular, note that doubling
the SFR in the last few Gyr does not introduce significant departures in the
WDLF. In addition, changes in the SFR at earlier times will be absorbed in the
normalization procedure of the theoretical WDLFs. Moreover, because we
restricted the normalization of the theoretical WDLFs and the comparison with
observations at $ M_{\rm Bol} \lesssim 13$, our comparison is restricted to
white dwarfs born in the last $\sim 1$ Gyr, which means that it is only
sensitive to fluctuations in the SFR in that period. In particular, it is
worth noting that hypothetical bursts in the SFR at very late times, that is,
shorter than 1 Gyr ago, as those suggested in \citet{2013MNRAS.434.1549R},
would produce an upward shift of the WDLF at high luminosities (relative to
the normalization luminosity range) and cannot be confused with an additional
cooling mechanism.  However, a burst in the SFR in the last $\sim 1$ Gyr might
help to hide the impact of extra cooling mechanisms in the WDLF. In view of
the previous discussion, a disk age of 11 Gyr was assumed throughout under the
assumption of a constant SFR (i.e., $T_D=11$Gyr).

\section{The WDLF of the Galactic disk}
\label{Sec:UniWDLF}
\begin{figure*}[ht!]
\begin{center}
\includegraphics[clip, angle=0, width=12cm]{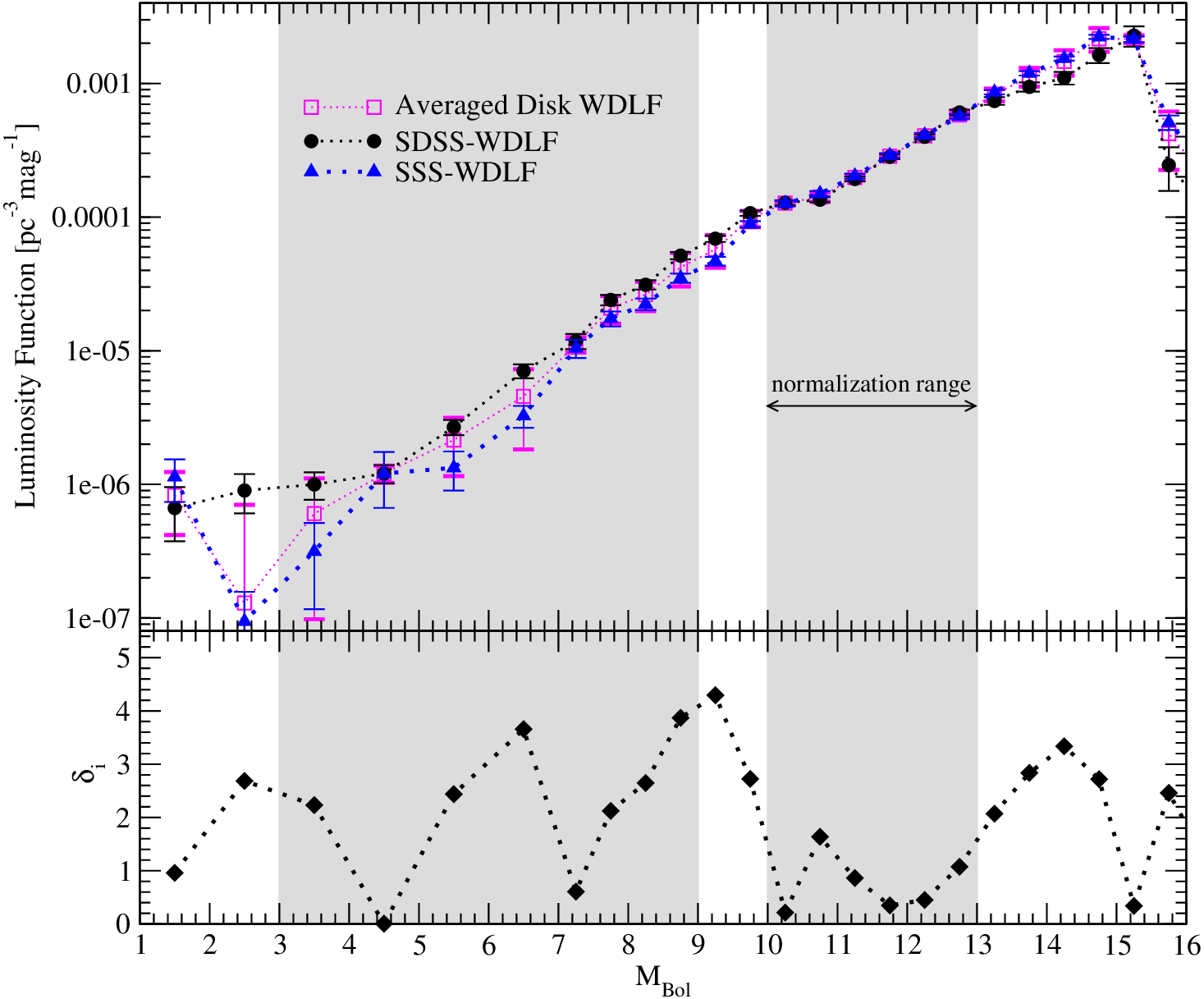} 
\caption{{\it Top panel:} Comparison between the SDSS- and
  SSS-WDLFs. Absolute values of the SSS-WDLF have been corrected for
  incompleteness by normalization to the SDSS-WDLF total number of
  stars in the range 10$<M_{\rm Bol}<$13 (small shaded area). Open
  squares show the averaged WDLF constructed by merging the WDLFs
  derived by \citet{2006AJ....131..571H}, \citet{2009A&A...508..339K},
  and \citet{2011MNRAS.417...93R}, see text for details. The gray area
  within 3$<M_{\rm Bol}<$9 corresponds to the region used for the
  $\chi^2$-like test. {\it Bottom panel:} Differences between the
  SDSS- and SSS-WDLF relative to their own quoted error bars at each
  magnitude bin ($\delta_i=|n_{i,{\rm SDSS}}-n_{i,{\rm
      SSS}}|/\sqrt{{\sigma_{i, {\rm SDSS}}}^2+{\sigma_{i,{\rm
          SSS}}}^2}$). Values of $\delta_i>1$ indicate significant
  discrepancies between the SDSS-WDLF and the SSS-WDLF.}
\label{Fig:WDLF_err}
\end{center}
\end{figure*} 
The SDSS has increased the number of known white dwarfs by more than one order
of magnitude in the past decade. Taking advantage of these large amounts of
data, several works have derived WDLFs in different magnitude ranges by means
of different techniques. \citet{2006AJ....131..571H} derived a WDLF from the
SDSS DR3 using the reduced proper motions technique for all white dwarfs
(without separating them into H-rich, DA, or H-deficient DB\footnote{While hot
  H-deficient white dwarfs are classified either as DO or DB, depending on
  their temperature, here we used the abbreviation DB to refer to all
  H-deficient white dwarfs.}). Because intrinsically bright WD stars can be
seen at much greater distances, on average they do not show large
proper motions and are not suited to the reduced proper-motion
technique. Thus, their WDLF has to be limited to relatively low
luminosities (7$<M_{\rm Bol}<$16). Using the same technique, but based on the
SDSS DR4 and constraining it solely to spectroscopically derived DA-WDs,
\citet{2008AJ....135....1D} derived a DA-only WDLF in the range 5.2$<M_{\rm
  Bol}<$12.4.
Also from the SDSS-DR4, but based on the color-selection technique,
which works well at high luminosities, \citet{2009A&A...508..339K}
derived high-luminosity WDLFs (both a H-rich-only WDLF and a
H-deficient WDLF) for the range 0$<M_{\rm Bol}<$7. Using the WDLFs of
\citet{2006AJ....131..571H} and \citet{2009A&A...508..339K}, we 
constructed a WDLF for DA+DB white dwarfs (from now on SDSS-WDLF).

All the WDLFs mentioned in the previous paragraph have been derived from the
SDSS catalog and might be prone to the same unknown systematic errors or
biases. In particular, because a proper assessment of the uncertainties is
crucial for the objective of the present work, we also included in our
analysis the WDLF derived from the SuperCosmos Sky Survey (SSS,
\citealt{2011MNRAS.417...93R}).  \citet{2011MNRAS.417...93R} measured the WDLF
from a sample of around 10000 WDs using the proper-motion technique to derive
a very deep WDLF, in the range $1<M_{\rm Bol}<18$. While the WDLF derived from
SSS is admittedly incomplete at around a 50\% level, it covers a larger area
of the sky and might be prone to different unknown systematic errors than the
SDSS data (see \citealt{2011MNRAS.417...93R, 2013MNRAS.434.1549R} for
details). In particular, it is claimed that the SSS incompleteness is uniform
and does not bias the WDLF \citep{2013MNRAS.434.1549R}. While
\citet{2011MNRAS.417...93R} developed a new method to derive their WDLF, we
here relied on their WDLF derived with the standard $V_{\rm max}^{-1}$
technique for more direct comparison with the SDSS data.  This allowed us to
perform a comparison of different, completely independent WDLFs, and make a
better assessment of the uncertainties behind the WDLF of the Galactic disk.

\subsection{Comparison between the SDSS and SSS WDLFs}
\label{Sect:comp}
A preliminary comparison between our theoretical WDLFs and the
SDSS and the SSS WDLFs suggested that the two WDLFs were discrepant
beyond their quoted error bars. To obtain a quantitative
measure of the differences between the two WDLFs we performed a
$\chi^2$-like statistical test. Because the WDLF derived from the SSS is
incomplete at around a 50\% level, absolute numbers of both WDLFs will
differ. Therefore, the WDLF from the SSS survey must first be rescaled
before one can compare it with the SDSS-WDLF.  Consequently, we rescaled the
WDLF of \citet{2011MNRAS.417...93R} so that the total number of stars
per unit volume in the range $10 \leq M_{\rm Bol} \leq 13 $ is similar
to the one in the SDSS-WDLF \citep{2006AJ....131..571H}, as we did
with the theoretical WDLFs. The derived correction factor is
$c=1.862$, which is consistent with the claimed incompleteness of
around 50\%. 

Then we redefined the values of the
\citet{2011MNRAS.417...93R} WDLF (RH11) in each magnitude bin as
$n_{i, {\rm SSS}}=c\,n_{i,{\rm RH11}}$ and $\sigma_{i, {\rm
    SSS}}=c\,\sigma_{i, {\rm RH11}}$. To define both WDLFs
at the same magnitude bins in the range $M_{\rm Bol}<7$, where
the WDLF of \citet{2009A&A...508..339K} is given at half magnitude
bins, we added the two values of the corresponding two bins of the
\citet{2011MNRAS.417...93R} as
$n_{i, {\rm
  SSS}}^{0.5}=(n_{i, {\rm SSS}}^{0.25}+n_{i, {\rm
  SSS}}^{0.75})/2$ 
and 
${\sigma_{i, {\rm SSS}}^{0.5}}^2=({\sigma_{i, {\rm
      SSS}}^{0.25}}^2+{\sigma_{i, {\rm SSS}}^{0.75}}^2)/4$ (from now
the SSS-WDLF).  This left two WDLFs given at the same
magnitude points and with the same number of stars per volume in the
magnitude range $10<M_{\rm Bol}<13$. Fig. \ref{Fig:WDLF_err} shows the
absolute and relative differences between the number densities of the
SDSS-WDLF and the SSS-WDLF relative to a measure of their quoted error
bars.

A more quantitative indication of the differences between the two WDLFs
can be obtained as follows: if both WDLFs were realizations of
distributions with the same mean value $\mu_i$ in each magnitude bin
and with the quoted variances\footnote{And under the standard
  assumption of Gaussian errors.}, then at each magnitude bin the
quantity $(n_{i,{\rm SDSS}}-n_{i,{\rm SSS}})/\sqrt{{\sigma_{i, {\rm
        SDSS}}}^2+{\sigma_{i,{\rm SSS}}}^2}$ would be a random
variable with a normal distribution of unit variance and zero
mean. Then,
\begin{equation}
\chi^2=\sum_{3<M_{\rm Bol, i}<9} \frac{(n_{i,{\rm  SDSS}}-n_{i,{\rm  SSS}})^2}
{{\sigma_{i,{\rm  SDSS}}}^2+{\sigma_{i,{\rm  SSS}}}^2}
\end{equation}
should follow a $\chi^2$ square distribution with eight degrees of
freedom. The probability that the actual value of $\chi^2=51.16$
occurs under the previous assumptions is lower than $P=10^{-7}$.

Clearly, the SDSS and SSS are not consistent within their quoted error bars.
Either some mean values are inaccurate beyond the quoted error bars, or error
bars in the WDLFs have been significantly underestimated\footnote{There is
  also a third possibility that the errors are highly non-Gaussian.}.  It is
worth noting that differences between SSS and SDSS WDLFs are not restricted to
high luminosities, where the proper-motion technique adopted by
\citet{2011MNRAS.417...93R} might not be best suited. In particular,
differences between \citet{2006AJ....131..571H} and
\citet{2011MNRAS.417...93R} WDLFs are very significant (in terms of their
quoted error bars) in the range 7.5$<M_{\rm Bol}<$9.5 (see
Fig. \ref{Fig:WDLF_err}). From this comparison we are forced to conclude that
at least one of the WDLFs is more uncertain than quoted by its own error
bars. Owing to the incompleteness of the SSS-WDLF, and the lack of reddening
correction, one possible cause for the observed inconsistency is that
incompleteness is not uniform at all magnitude bins, which biases the final
SSS-WDLF (Rowell, private communication).  In addition, the lack of reddening
corrections in the SSS-WDLF would tend to affect bright magnitudes more, which
would also bias the SSS-WDLF. The other possible cause for this discrepancy is
just that uncertainties are larger than quoted in both WDLFs.  In view of this
discrepancy and in the absence of a third independent WDLF we decided on a
two-way approach. On the one hand, we assumed the two WDLFs to be equally
valid and derived from them an averaged WDLF (Sect. \ref{Sect:Uni-WDLF}). This
averaged WDLF was then compared with theoretical models to obtain constraints
for $\mu_\nu$ (Sect. \ref{Sect:Mu}). On the other hand, we assumed that the
inconsistency is due to some unexplained bias in the SSS-WDLF and compared the
theoretical models directly with the SDSS-WDLF (Sect. \ref{Sect:Mu}).

In the next section we estimate a WDLF taking into account the
systematic differences between the two WDLFs.

\subsection{Averaged WDLF of the Galactic Disk}
\label{Sect:Uni-WDLF}
\begin{figure*}[ht!]
\begin{center}
\includegraphics[clip, angle=0, width=12cm]{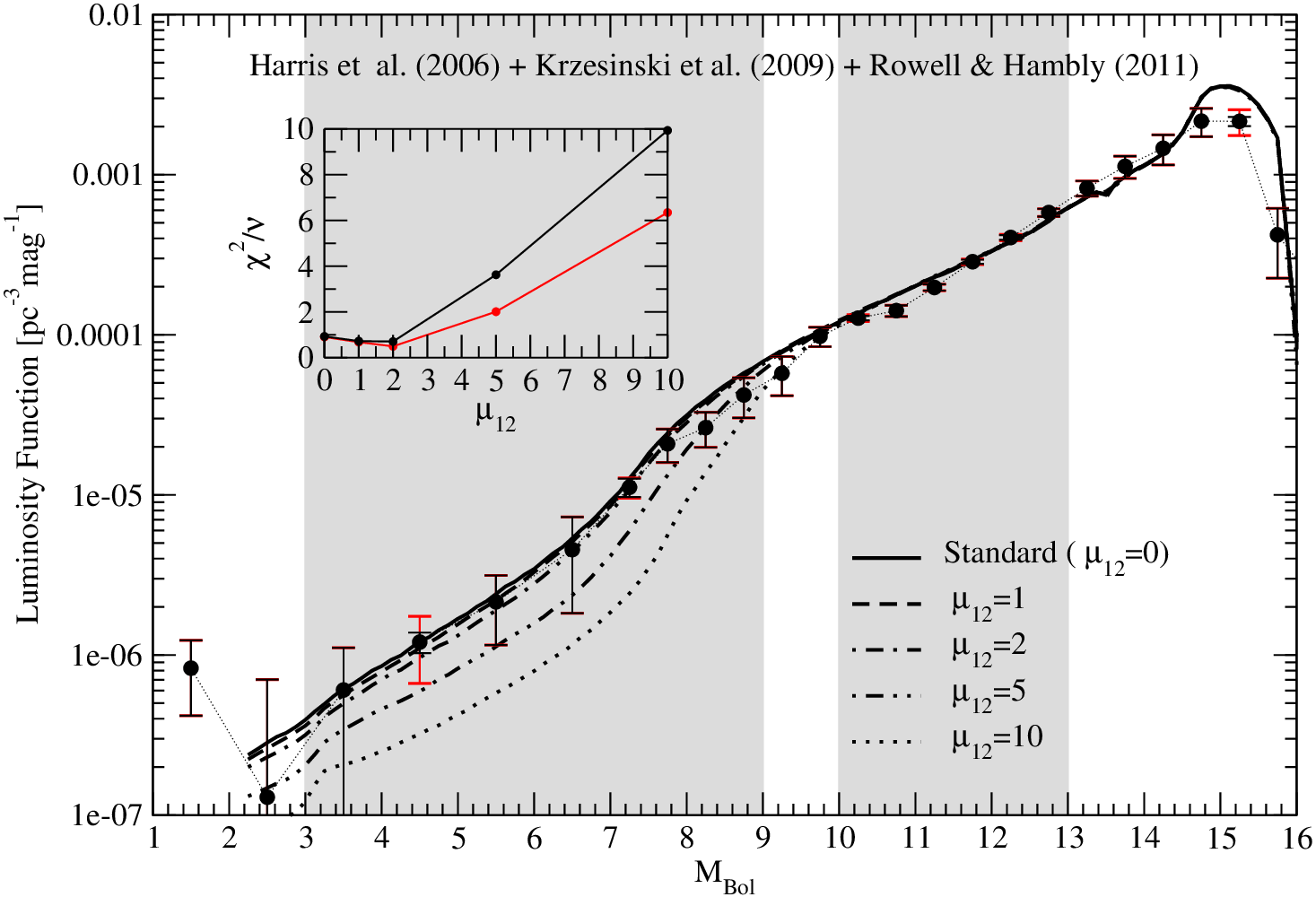} 
\caption{Comparison of our theoretical WDLFs constructed under the
  assumption of different values of $\mu_{12}$ with the unified WDLF
  constructed merging the WDLFs derived by
  \citet{2006AJ....131..571H}, \citet{2009A&A...508..339K}, and
  \citet{2011MNRAS.417...93R}. Red error bars correspond to those
  derived as $\sigma_i={\rm Max}[\sigma_{i, {\rm Uni}},\sigma_{i, {\rm
      SSS}},\sigma_{i, {\rm SDSS}}]$. See Sect. \ref{Sec:UniWDLF} for
  details. The inset shows the value of the $\chi^2$ per degree of
  freedom $\nu$ of the $\chi^2$-test for the two different sets of
  error bars. Gray areas indicate the magnitude ranges used for the
  $\chi^2$-test (3$<M_{\rm Bol}<$9) and for the normalization
  procedure 10$<M_{\rm Bol}<$13. }
\label{Fig:WDLF_Uni}
\end{center}
\end{figure*}
The comparison between the SDSS- and SSS- WDLFs suggests that the
uncertainties in the WDLFs have been significantly underestimated. This led us
to use both the SDSS and SSS to obtain a unified WDLFs with error bars that
take into account the differences between the two derived WDLFs.  If the
values $n_{i, {\rm SSS}}$ and $n_{i,{\rm SDSS}}$ were consistent in each
magnitude bin, then it is clear that the best estimation of the WDLF would be
a weighted average of the values and the new uncertainties would be given by
$\sigma_i^{-2}={\sigma_i^{\rm SSS}}^{-2}+{\sigma_i^{\rm SDSS}}^{-2}$. However,
the uncertainty estimated in this way would not reflect the differences in the
values of ${\bar{n_i}}^{\rm SSS}$ and ${\bar{n_i}}^{\rm SDSS}$, as it
should. In fact, dealing with discrepant data is difficult (see for
example \citealt{2005AIPC..803..431H}). A detailed systematic treatment of the
discrepant SSS and SDSS data is beyond the scope of the present
work. However, an intuitively reasonable estimation of the unified WDLFs and
its uncertainties is given by
\begin{equation}
n_{i,{\rm Uni}}=\frac{\sigma_{i,{\rm SSS}}^{-2}n_{i,{\rm SSS}}+\sigma_{i,{\rm
SDSS}}^{-2}n_{i, {\rm SDSS}}} {\sigma_{i,{\rm SSS}}^{-2}+\sigma_{i,{\rm
    SDSS}}^{-2}}
\label{eq:mean}
\end{equation}
and
\begin{equation}
\sigma_{i, {\rm Uni}}^2=\frac{1}{\sigma_{i, {\rm SSS}}^{-2}+\sigma_{i,{\rm
    SDSS}}^{-2}}+\frac{(n_{i,{\rm SSS}}-n_{i, {\rm SDSS}})^2}{2}.
\label{eq:sig}
\end{equation}
Note that this expression for $\sigma_i^{\rm Uni}$ is restricted to the
standard variance summation rule when ${\bar{n_i}}^{\rm SSS}\sim{\bar{n_i}}^{\rm
  SDSS}$, and for significantly discrepant data it
approaches the value of the unbiased sample variance estimator (for a
two-point sample). Thus, this estimation of the uncertainties in the
WDLFs has the advantage that it reduces the uncertainty in those bins
where the SDSS and SSS WDLFs are consistent and increases the
uncertainty when the SDSS and SSS WDLFs are discrepant.
\begin{table}
{\small
\caption {Averaged WDLF of the Galactic disk} 
\label{tab:disk-WDLF} 
\begin{center}
   \begin{tabular}{ c | c | c | c } 
     M$_{\rm Bol}$ & $n_{{\rm Uni}}$ & $\sigma_{{\rm Uni}}$ &${\rm
       Max}{[\sigma_{{\rm  SDSS}},\,\sigma_{{\rm  SSS}},\,\sigma_{{\rm
             Uni}}]}$ \\ 
& [pc$^{-3}$mag$^{-1}$]& [pc$^{-3}$mag$^{-1}$] & [pc$^{-3}$mag$^{-1}$] \\\hline
   1.50   &     8.28$\times 10^{-7}$ &  4.10$\times 10^{-7}$ & 4.10$\times 10^{-7}$  \\ 
   2.50   &     1.30$\times 10^{-7}$ &  5.73$\times 10^{-7}$ & 5.73$\times 10^{-7}$  \\ 
   3.50   &     6.04$\times 10^{-7}$ &  5.07$\times 10^{-7}$ & 5.07$\times 10^{-7}$  \\ 
   4.50   &     1.20$\times 10^{-6}$ &  1.77$\times 10^{-7}$ & 5.40$\times 10^{-7}$  \\ 
   5.50   &     2.15$\times 10^{-6}$ &  9.95$\times 10^{-7}$ & 9.95$\times 10^{-7}$  \\ 
   6.50   &     4.56$\times 10^{-6}$ &  2.73$\times 10^{-6}$ & 2.73$\times 10^{-6}$  \\ 
   7.25   &     1.12$\times 10^{-5}$ &  1.47$\times 10^{-6}$ & 1.62$\times 10^{-6}$  \\ 
   7.75   &     2.09$\times 10^{-5}$ &  4.90$\times 10^{-6}$ & 4.90$\times 10^{-6}$  \\ 
   8.25   &     2.63$\times 10^{-5}$ &  6.48$\times 10^{-6}$ & 6.48$\times 10^{-6}$  \\ 
   8.75   &     4.21$\times 10^{-5}$ &  1.19$\times 10^{-5}$ & 1.19$\times 10^{-5}$  \\ 
   9.25   &     5.75$\times 10^{-5}$ &  1.58$\times 10^{-5}$ & 1.58$\times 10^{-5}$  \\ 
   9.75   &     9.78$\times 10^{-5}$ &  1.36$\times 10^{-5}$ & 1.36$\times 10^{-5}$  \\ 
   10.25  &     1.27$\times 10^{-4}$ &  4.26$\times 10^{-6}$ & 6.14$\times 10^{-6}$  \\ 
   10.75  &     1.41$\times 10^{-4}$ &  1.13$\times 10^{-5}$ & 1.13$\times 10^{-5}$  \\ 
   11.25  &     1.98$\times 10^{-4}$ &  9.10$\times 10^{-6}$ & 9.10$\times 10^{-6}$  \\ 
   11.75  &     2.86$\times 10^{-4}$ &  8.90$\times 10^{-6}$ & 1.16$\times 10^{-5}$  \\ 
   12.25  &     4.05$\times 10^{-4}$ &  1.38$\times 10^{-5}$ & 1.80$\times 10^{-5}$  \\ 
   12.75  &     5.80$\times 10^{-4}$ &  3.24$\times 10^{-5}$ & 3.24$\times 10^{-5}$  \\ 
   13.25  &     8.25$\times 10^{-4}$ &  8.90$\times 10^{-5}$ & 8.90$\times 10^{-5}$  \\ 
   13.75  &     1.13$\times 10^{-3}$ &  1.79$\times 10^{-4}$ & 1.79$\times 10^{-4}$  \\ 
   14.25  &     1.46$\times 10^{-3}$ &  3.12$\times 10^{-4}$ & 3.12$\times 10^{-4}$  \\ 
   14.75  &     2.16$\times 10^{-3}$ &  4.34$\times 10^{-4}$ & 4.34$\times 10^{-4}$  \\ 
   15.25  &     2.15$\times 10^{-3}$ &  1.43$\times 10^{-4}$ & 3.95$\times 10^{-4}$  \\ 
   15.75  &     4.21$\times 10^{-4}$ &  1.96$\times 10^{-4}$ & 1.96$\times 10^{-4}$  \\ 
   16.25  &     2.09$\times 10^{-4}$ &  1.09$\times 10^{-4}$ & 1.09$\times 10^{-4}$  \\ 
   \end{tabular}
\end{center}
}
\end{table}

The unified disk WDLFs derived in this way (Table \ref{tab:disk-WDLF})
can now be compared with the theoretical WDLFs. For the objective of
the present work, the most significant difference between this new
WDLF and the SDSS and SSS WDLFs is that the error bars now reflect the
discrepancies between different estimations of the WDLFs.

\section{Constraints on $\mu_\nu$ and discussion}
\label{Sect:Mu}

\begin{figure}
\includegraphics[clip, angle=0, width=8.5cm]{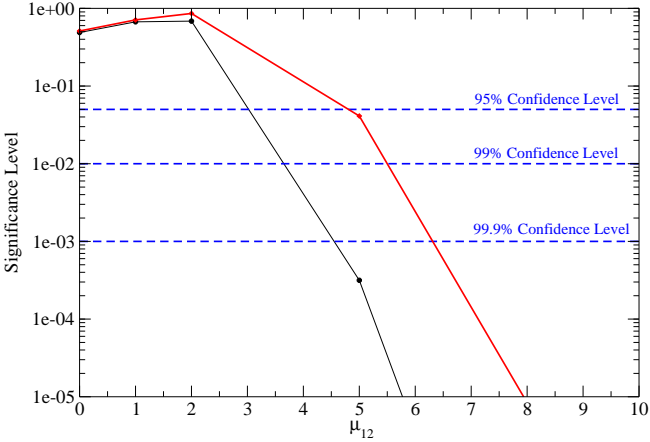} 
\caption{Significance level of the $\chi^2$-test for the WDLFs shown
  in Fig. \ref{Fig:WDLF_Uni}. It is clear that, under the error
  estimation presented in Eq. \ref{eq:sig}, magnetic dipole moments
  larger than $\mu_{12}= 5$ can be discarded at the 99.9\% confidence
  level. Black (red) line shows the result when the value of
  $\sigma_{{\rm Uni}}$ (${\rm Max}{[\sigma_{{\rm
          SDSS}},\,\sigma_{{\rm SSS}},\,\sigma_{{\rm Uni}}]}$) is
  adopted as the uncertainty ---see Table \ref{tab:disk-WDLF}. }
\label{Fig:chi2}
\end{figure}

In Fig. \ref{Fig:WDLF_Uni} we compare the WDLF derived in the previous section
by averaging both the SDSS- and SSS-WDLFs with the theoretically computed ones
under the assumption of different values of the magnetic dipole moment of the
neutrino. In addition, in Fig. \ref{Fig:chi2} we show the result of the
$\chi^2$-test performed on the averaged disk-WDLF derived in
Eqs. \ref{eq:mean} and \ref{eq:sig}. As can be directly appreciated from
Figs. \ref{Fig:WDLF_Uni} and \ref{Fig:chi2}, the $\chi^2$-test shows that
values of $\mu_{12}\gtrsim 5$ are significantly at variance with the
observations. In fact, when adopting the error bars derived in
Eq. \ref{eq:sig} (black line in Figs. \ref{Fig:WDLF_Uni} and \ref{Fig:chi2}),
Fig. \ref{Fig:chi2} shows that WDLFs constructed with $\mu_{12}\ge 5$ can be
rejected at more that a 99.9\% confidence level (i.e. $\gtrsim
3\sigma$-like). Even when a more conservative error estimation is adopted
(${\rm Max}{[\sigma_{{\rm SDSS}},\,\sigma_{{\rm SSS}},\,\sigma_{{\rm Uni}}]}$,
red lines in Fig. \ref{Fig:WDLF_Uni} and \ref{Fig:chi2}), values of
$\mu_{12}\ge 5$ can be rejected at more that a 95\% confidence level
(i.e. $\sim 2\sigma$-like).

\begin{figure*}[t!]
\begin{center}
\includegraphics[clip, angle=0, width=12cm]{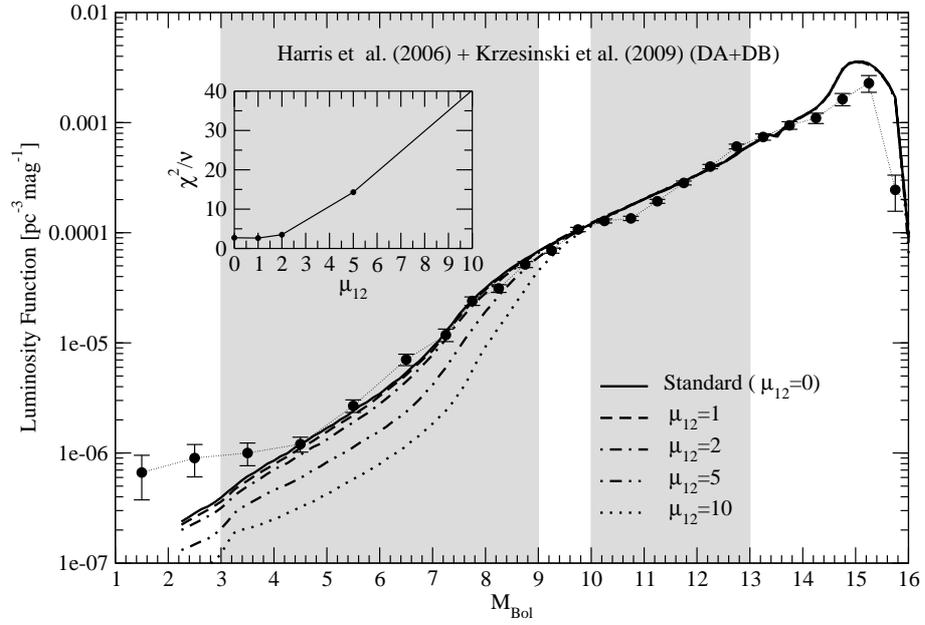} 
\caption{Comparison of our theoretical WDLFs constructed under the
  assumption of different values of $\mu_{12}$ with the SDSS-WDLF (for
  all WDs) derived by \citet{2006AJ....131..571H} and
  \citet{2009A&A...508..339K}. The inset shows the value of the
  $\chi^2$ per degree of freedom $\nu$ of the $\chi^2$-test. Gray
  areas indicate the magnitude ranges used for the $\chi^2$-test (3$<M_{\rm Bol}<$9) and for the normalization procedure 10$<M_{\rm Bol}<$13.}
\label{Fig:WDLF_Harris}
\end{center}
\end{figure*}

As noted in Sect. \ref{Sect:comp}, the discrepancy between the SDSS- and
SSS-WDLFs might also be related to some unexplained bias (e.g. incompleteness)
in the SSS-WDLF. If this is the case, the SDSS-WDLF should be preferred. In
Fig. \ref{Fig:WDLF_Harris} we compare the theoretical WDLFs with the
SDSS-WDLF, also by means of a $\chi^2$-test. It is clear from this comparison
that values of $\mu_{12}\ge 5$ are at variance with observations. However, in
this case, the $\chi^2$-values are still too high for $\mu_{12}=$0, 1 and 5
and the models do not fit the observations because the error bars are
significantly smaller than in the averaged WDLF.  The failure of the models to
fulfill a $\chi^2$-test can be because either the quoted error
bars are too low (as suggested by the comparison with the SSS-WDLF) or
because the uncertainties in the theoretical models become more relevant. In
the latter case the impact of the possible existence of short-term
fluctuations in the SFR should be explored.

It is worth noting that our theoretical WDLFs are derived only taking into
account DA-WD models. This would naively suggest that our theoretical WDLF
should be compared with observational WDLFs only for DAs
\citep{2008AJ....135....1D}. While this might in principle be true, it might
also be misleading because of our current lack of a complete understanding of
the spectral evolution of white dwarfs ---see \citet{1987fbs..conf..319F} for
an early description of the problem. In particular, white dwarfs with an
extremely thin H-envelope such as those studied by
\citet{2013EPJWC..4305006S}, and references therein, will cool as DB white
dwarfs, but will still be classified as DA. Even worse, the fraction of such
DA white dwarfs with extremely thin H-envelopes probably depends on the
magnitude bin. In view of this situation, we chose to compare our theoretical
WDLF with DA-only and DA+DB WDLFs. Fortunately, differences between DA-only
and DA+DB WDLFs are most likely on the order of 10 to 20\%
\citep{2008AJ....135....1D}, values which are within the error bars of our
averaged WDLF (Table \ref{tab:disk-WDLF}). Indeed, when a $\chi^2$-test was
performed on the individual WDLFs derived from the SDSS, they all yielded
similar conclusions, regardless of whether they were DA-only WDLFs
\citep{2008AJ....135....1D}%
\footnote{When comparing the WDLF of \cite{2008AJ....135....1D} with
  theoretical models, we restricted the normalization region and the range of
  the $\chi^2$-test to the regions were it is valid (10$<M_{\rm Bol}<$12.4 and
  5.2$<M_{\rm Bol}<$9.2, respectively).}
or DA+DB WDLFs \citep{2006AJ....131..571H, 2009A&A...508..339K}. In short, the
comparison of our theoretical WDLFs with all the WDLFs derived from the SDSS
suggest that models with $\mu_{12}<5$ agree better with the observed disk
WDLFs. However, because of the significantly lower error bars, a $\chi^2$-test
indicates that all theoretical models fail to fit the observations within the
quoted error bars.  The same is true for a $\chi^2$-test performed on the
WDLF derived from the SSS, because the low error bars the test rejects all
theoretical models ---although in this case theoretical models with
$\mu_{12}=5$ and 10 are closer to the observations. We consider this to be
another indication that the quoted error bars in those works might be
underestimated, although this result can also be related to the absence of an
estimation of the systematic errors in theoretical models. Certainly, this
result calls for a larger set of completely independent WDLFs as well as for
the derivation of WDLFs by means of different WDLF estimators
\citep{2006MNRAS.369.1654G}.

\section{Summary and conclusions}

We have compared the observed WDLFs derived from two different surveys (SDSS
and SSS) and showed that these WDLFs are not consistent within their quoted
error bars. Consequently, we constructed a unified disk-WDLF (Table
\ref{tab:disk-WDLF}) by averaging the SSS and SDSS WDLFs and estimating the
uncertainty in the derived values by taking into account the discrepancy
between the two sets of data as well as their own quoted error bars. Then we
used this averaged disk-WDLF to constrain neutrino physics. To this aim we
computed 25 white dwarf evolutionary sequences under the assumption of
different values of the neutrino magnetic dipole moment ($\mu_{12}=$0, 1, 2,
5, and 10). With these sequences, theoretical WDLFs for the Galactic disk were
computed for the different values of $\mu_{12}$ and compared with the
observations. A $\chi^2$-test on the unified disk-WDLF (SDSS+SSS) yielded that
values of $\mu_{12}\geq 5$ can be rejected at more than a 95\% confidence
level, even when a conservative estimation of the error bars is
adopted. Moreover, a direct comparison with the SDSS-WDLFs suggested that
values of $\mu_{12}> 5$ can be rejected. This result is not far from the best
available constraints on $\mu_{12}$ from the CMD of globular clusters and is
based on independent astronomical determinations.  This results shows the
power of the new WDLFs to constrain the value of the magnetic dipole moment of
the neutrino. However, the discrepancy between the SDSS and SSS WDLFs needs to
be addressed, probably by means of a larger set of completely independent
WDLFs.

Future determinations of the WDLF of the Galactic disk based on independent
surveys  and different WDLF estimators will allow a better determination
of the actual disk-WDLF. In addition to a better statistical treatment of all
available WDLFs, other improvements should be performed in the computation of
the theoretical WDLFs. First, to improve the constraints derived in
this work it would be desirable to construct WDLFs that include the
contribution of H-rich and H-deficient white dwarfs. A systematic
exploration of the impact of the uncertainties in the SFR of the disk in the
past Gyr needs to be made to estimate systematic errors in the
  comparison of theoretical and inferred WDLFs.  Finally, it would be
interesting to test the impact of a magnetic dipole moment on the pre-white
dwarf stages and how they affect the derived white dwarf models. We believe
that, once this is done, the WDLF will offer constraints on the magnetic dipole
moment of the neutrino similar to the best available constraints obtained from
the CMD of globular clusters.

\begin{acknowledgements}
     M3B is supported by a fellowship for postdoctoral researchers from the
     Alexander von Humboldt Foundation. M3B thanks H. Harris, N. Rowell,
     N. Hambly, S. DeGennaro and J. Krzesinski for the data and instructive
     comments about their respectively WDLFs, J. Isern, G. Raffelt and
     N. Rowell for reading and commenting on a preliminary version of the
     article, and J. Beacom for comments and suggestions that have strongly
     improved the final version of the manuscript. L. Althaus is warmly
     thanked for extensive and instructive discussions about the physics of
     white dwarfs throughout the years. This research was partially supported
     by PIP 112-200801-00940 from CONICET and by ANPCyT through PMT III (BID
     1728/OCAR). This work is dedicated to the memory of my uncle, Miguel
     Bertolami, who introduced me to science.
\end{acknowledgements}

\bibliographystyle{aa}
\bibliography{mmiller}

\end{document}